# Comment on "Propagation and Negative Refraction", IEEE Microwave Magazine, pp. 58-65, July/August 2012.

Hunter et al. have recently presented a confusing interpretation of the phenomenon of negative refraction in metamaterials [1]. In the first part of their article, the authors make bold and general claims that negative refraction is not possible at a single interface due to causality issues. They attempt to explain the decade-old experimental and numerical evidence of negative refraction with a couple of paragraphs on the premise that this is only possible in the case of a finite-thickness slab based on modal interference. To support their claim, the authors continue by proposing an even/odd modal analysis which would presumably explain all the observed negative refraction phenomena. They then go on applying this theory to lumped-circuit networks and to specific metamaterial geometries that have successfully reported negative refraction based on a variety of different phenomena. In this comment, we would like to clarify and rectify several misleading statements appearing in their article and straighten the physical meaning and underlying phenomena behind negative-refraction and negative-index metamaterials.

Central to their denial of negative refraction at a single interface is the apparent violation of Einstein's causality, as explained in the text accompanying Fig. 2 in [1]. This causality argument is not new and was first proposed in 2002 [2]. It appears that Hunter et al. are unaware of the subsequent debate around the claims in [2], which has been overwhelmingly resolved in favor of the conventional explanation of negative refraction, as described in the following. It should be first noticed that the schematic picture in Fig. 2 in [1] assumes an infinitely extent monochromatic plane-wave in the steady-state excitation. Under these conditions, i.e., with an excitation infinitely extended both in space and time, there is no causality issue when assuming negative refraction at a single interface, since the wave has been established since infinite time and occupies all the available space. In such case, causality cannot be invoked and what determines refraction is simply phase velocity, which can assume positive, negative, zero or infinite values without violating any physical law (of course, the energy velocity, which is in general different from phase velocity, is always positive and less than the vacuum speed of light). In order to argue about causality, one needs to consider pulses with a finite extent in space-time. However, as noticed in several papers in the last decade, when considering realistic excitations with finite spatial and temporal extent one should also take into account the spatial and temporal dispersion of the refraction phenomenon. For what concerns the temporal response, it is well known [3] that materials with bulk constitutive parameters smaller than those of free-space are necessarily frequency-dispersive and this implies that the index of refraction of the material on the right in Fig. 2 cannot be negative for all frequencies. For this reason, when considering an excitation starting at an instant in time, the initial part of a pulse hitting the interface of Fig. 2 would necessarily refract positively (to be precise, the precursor of any finite signal hitting the interface would not refract at all, as the very high-frequency response of any passive material is the same as free-space [3,4]). However, after a transient period, an applied narrow-band signal with its middle-band frequency centered in the negative-index band of the metamaterial at the right of Fig. 2 would indeed experience negative refraction at a single interface, contrary to the claims of Hunter et al. It is now very well-understood that the establishment of negative refraction requires a finite transient period before the power can refract negatively at a single interface

[5]. Beyond these theoretical explanations, in the literature there are countless examples of negative refraction at a single interface [6-8]. When considering beams with finite spatial extent, such as modulated waves, it has also been conclusively shown and it is well accepted that the group velocity and power flow refract negatively at a single interface as in Fig. 2, as proven in [9,10]. Therefore, evidently unbeknownst to Hunter et al., the main argument of their paper [1] has been totally dismissed in the literature since several years ago, and we are puzzled why this issue is still being considered. After their supposed 'proof' that negative refraction is impossible at a single interface, the authors go on proposing a strange model to explain the phenomenon of negative refraction based on an even/odd modal analysis of coupled lines. Based on this model, they attempt at explaining what they assume to be negative refraction through a finite metamaterial slab, as shown in Fig. 3. *However, Fig. 3 does not describe negative refraction at all*! It is well known and obvious that at the exit of any finite-thickness slab, possessing a negative refractive index, the wave must undergo two negative refractions (entrance and exit interfaces) and therefore must emerge parallel with the original incidence angle, but just laterally shifted in the 'negative' direction as compared to the original incident wave. This process may also be easily understood using just a simple ray picture, as originally envisioned in Veselago's seminal paper describing negative-index materials [11].

All the experimental and numerical results reported on finite-thickness negative-index metamaterials to date have reported a transmitted beam that emerges parallel with the incident beam but laterally negatively shifted due to the double negative-refraction at the entrance and exit faces of the negative-index slab. This also follows from momentum conservation and phase matching. In Fig.3 (and Fig. 8, and Fig. 18, and Fig. 20) of [Hunter], the angles $\theta_i$ and $\theta_r$ have opposite signs, and thus phase matching is violated and momentum is not conserved, so what the authors are trying to achieve is not consistent with any transversely homogeneous material slab. The authors seem to be aiming at *phase conjugation* at the exit of a finite-thickness slab, which is very different from negative-refraction or negative-index metamaterials. The phase reversal at the slab exit is central to the analysis that Hunter et al. propose, as described mathematically in equation (6). However, none of the references 1,2,4,5 in [1] claim such kind of phenomenon, which is very distinct from negative-index metamaterials. The reader is referred to Figure 6 in [6] and Figure 10 in [7] for the correct picture of negative refraction through a finite-thickness metamaterial slab. It should be noted that, in these and many other examples available in the literature, negative refraction does not depend on a special condition on the slab thickness to satisfy the assumed phase conjugation condition (6), nor on the proper interference between two distinct eigenmodes, as in the model proposed in [1]. We can therefore readily conclude that the rest of the paper [1] is irrelevant to the issue of negative refraction and does not merit further consideration. It should be pointed out anyhow that, contrary to the claim in [1], the circuit analysis of the 2D transmission-line metamaterials in references 1 and 4 (Chapter 3) of [1] is complete and includes the use of *rigorous* Bloch-Floquet/[ABCD]-matrix analysis, and 2D Agilent-ADS circuit simulations based on the complete application of Kirchhoff's laws. Further details on the analysis of negative-refraction at a single interface between two transmission-line metamaterials can be found in [12]. More recently, the compatibility of these earlier published results with a complete multimodal circuit analysis has been thoroughly established in [8].

In addition, just to set the record straight, there are few other misleading statements and discussions that require clarification. The authors claim that the wire medium considered in [13] is an example applicable to their model. We doubt this assertion, as the wire medium is known to be well homogenizable as a bulk material in the long-wavelength limit [14]. Furthermore, we would like to clarify that the geometry presented in [13] is an example of negative-refracting material *without* negative *index* of refraction. In this specific geometry, negative refraction at an interface with air is obtained due to the hyperbolic anisotropy of the material, rather than to a negative-index isotropic medium, which is implicitly assumed in all the previous discussions in [1]. For this reason, this example presented by the authors is misleading and not applicable to the previous discussions. Moreover, Hunter et al. appear to have *used a TE incident wave* in their wire-medium analysis *which cannot* lead to negative refraction since in their case there is no electric field component along the wires, which was essential in the demonstration of negative refraction in [13]. Subsequently, the authors of [1] propose a flawed model for a metamaterial prism (Fig. 19) composed of cascaded coupled-line couplers. For their model to work the angle of incidence $\theta_i$ must be restricted to a single value related to the "prism angle" $\Phi$ in a certain way. In reference 9 of [1] though, the incident plane wave *is normal* to the input surface of the prism and negative refraction is observed at the slanted face of the prism (i.e. this is single-interface negative refraction)! This normal incidence excitation is typical in negative-refraction prism setups and it is relevant exactly to prove that negative refraction can happen at a single interface, contrary to all the erroneous claims in [1]; for another example of this phenomenon, the interested reader is referred to a paper by some of the makers of the popular EM fullwave solver CST [15].

In conclusion, the physics of negative refraction at a single interface between two homogeneous materials is well-established and it definitely does not violate Einstein's causality. What the authors of [1] appear to describe as refraction through a metamaterial slab is not refraction at all, but phase conjugation. A number of their other misrepresentations and confusing arguments have also been pointed out in this letter. Therefore, the entire claims in [1] can be easily refuted. It is hoped that this letter draws the attention of authors, reviewers and readers to the large body of rigorous analytical, numerical, and experimental evidence in favor of negative refraction at a single interface, so as to avoid reviving settled debates which mislead, rather than educate, the microwave community.

We would like to acknowledge Prof. Nader Engheta (U. of Pennsylvania), Prof. Xiang Zhang (U.C. Berkeley), Prof. Ashwin Iyer (U. Alberta), Dr. Rubaiyat Islam (AMD/Canada) and Prof. Mário Silveirinha (University of Coimbra, Portugal) for useful discussions.

**George V. Eleftheriades, U. Toronto, ECE**

**Andrea Alù, U. Texas at Austin, ECE**